\documentclass[reprint,
               superscriptaddress,
               nofootinbib,
               nobibnotes,
               bibnotes,
               amsmath,amssymb,
               aps,
               pra,
               floatfix,
              ]{revtex4-2}
\usepackage[utf8]{inputenc}
\usepackage{amssymb}
\usepackage{amsfonts}
\usepackage{amsmath}
\usepackage{cases}
\usepackage{stmaryrd}\usepackage{tipa}
\usepackage[dvips,pdftex]{graphicx}
\usepackage{dcolumn}
\usepackage{graphicx}
\usepackage{subfigure}
\usepackage{dcolumn}
\usepackage{bm}
\usepackage{color}
\usepackage{xcolor}
\usepackage{CJK}
\usepackage{physics}
\usepackage{array}
\usepackage{nopageno}
\usepackage{mathtools}
\usepackage[colorlinks=true, linkcolor=red, citecolor=blue, hyperfootnotes=false]{hyperref}

\DeclareMathOperator{\diff}{d}

\usepackage{braket}
\usepackage{esvect}
\usepackage{accents} 
\usepackage{ushort}
\usepackage{setspace}
\usepackage{afterpage}
\newcommand\partialb{\mathop{}\!{\bm{\partial}}}  
\newcommand{\metric}{\bm{g}}
\newcommand{\manifold}{\mathcal{M}}
\renewcommand{\vec}[1]{\mathbf{#1}} 
\newcommand*\spacetime{\mathop{}\!{(\mathcal{M},\bm{g})}}
\newtheorem{defn}{Definition} 

\begin{document}

\title{Diffeomorphism invariance and general covariance: a pedagogical introduction}

\author{Mateo Casariego}
\email{mateo.casariego@tecnico.ulisboa.pt}
\affiliation{Instituto Superior Técnico, Universidade de Lisboa, Portugal}
\affiliation{Physics of Information and Quantum Technologies Group, Centro de Física e Engenharia de Materiais Avançados (CeFEMA), Portugal}
\affiliation{PQI -- Portuguese Quantum Institute, Portugal}

\begin{abstract}
Diffeomorphism invariance is a feature that gets sometimes highlighted as something with profound implications in the physics of spacetime. Moreover, it is often wrongly associated exclusively with General Relativity. The fact that diffeomorphism invariance and general covariance are used interchangeably does not help. Here, we attempt at clarifying these concepts.
\end{abstract}
\maketitle

\section{Introduction}
Back in 2015, when what follows was written, it seemed to me that, in the context of General Relativity (GR), the notions of (general) covariance and diffeomorphism invariance were confusing to me. In the literature, an partly, perhaps, because of Einstein himself \cite{nortongencovar1995}, the terms were used as synonyms.
Here, we revisit our personal notes: an attempt at a self-contained discussion of the differences between the two, stressing the importance of active diffeomorphism invariance as one of the most unique features of GR.
It will be assumed that the reader has a fair level of familiarity with GR and differential geometry. Still, the aimed audience for this article is not the specialist. Rather, we hope that our exposition will help an undergraduate student understand -- and consequently admire --  the notion of diffeomorphism invariance and, more importantly, why it is something unique of GR. We apologize for the short and certainly not up to date references\footnote{After all, this is almost ten years old work!}. Good complementary references can be found in \cite{Barenz:2012av, RovelliGaulLQG&Diffeomorphism2000,  nortongcovariancegr1993, Norton2003-GeneralCovariance&Gauge,Pooley2015GRdiffeo, rickles, Rovelli:2013fga}.

\section{Tensors and notation}
Tensors are defined with respect to some group of transformations. A relation between tensors still holds if we transform them with respect to the group used to define them. In Minkowski spacetime this group is the Poincaré group\footnote{Boosts (in the Lorentz way), rotations, and translations.}: if we apply a Poincaré transformation to a Poincaré tensor, we will still have a Poincaré tensor. Contrary to elementary intuition, partial derivatives are not necessarily invariant operations, group-tensor-wise. In Minkowski spacetime we have a `Poincaré covariant derivative', which happens to be the same as the partial derivative\footnote{In coordinates such that $\mathrm{d} s^2=\eta_{\mu\nu}\mathrm{d} x^\mu \otimes \mathrm{d} x^\nu$.}; as long as we use it in our tensorial equations, their `covariance' is guaranteed. In GR, tensors are \textit{general}: they are defined to behave as tensor fields under \textit{any} (local) invertible coordinate transformation. The `general covariant derivative' (we will drop the `general' part) is the right derivative to use to preserve the tensorial character of our equations.

In a coordinate chart, the covariant derivative reads
\begin{equation}
{\bm{\nabla}}_{\bm{X}}\bm{t}= X^a\nabla_a\bm{t},
\end{equation}
where the subscript $a$ in $\nabla_a$ is a covariant \textit{index} rather than a vector field (it really means ${{\nabla}}_{\partialb_a}$). Defining the \textit{connection coefficients} $\Gamma^c_{\, ab}(x)$ as 
\begin{equation}
\Gamma^c_{\; ab}(x)\partialb_c \equiv \nabla_a (\partialb_b),
\end{equation}
we get the well-known expression for the covariant derivative along the coordinate axis $\partialb_a$ of the vector field $\bm{V}$:
\begin{equation}
{{\nabla}}_a\bm{V}=\left((\partial_aV^c)+\Gamma^c_{\; ab}V^b\right)\partialb_c.
\end{equation}

We will assume a Levi-Civita connection, and denote covariant differentiation of tensor $\bm t$ along the basis vector field $\partialb_a$ as $\bm{t}_{;a}$. For example, for a scalar field $\phi$ we have $\nabla_a \nabla^a \phi = \phi_{;a}^{\,\,\, ;a}.$

\begin{defn}\label{defn:spacetime}
A \textit{spacetime} consists on the pair $\spacetime$, where $\mathcal{M}$ is a four-dimensional real smooth (\emph{i.e.},	$\mathcal{C}^\infty$) manifold equipped with a Lorentzian metric\footnote{See below for a definition of a metric as a mapping.} $\metric$. 
\end{defn}

\section{General covariance and diffeomorphism invariance}

Our (simplified) principle of general covariance can be stated as follows: 
\begin{quote}
{\setstretch{1.0}
\emph{A theory is} generally covariant \textit{iff every equation describing its `laws' looks the same in all frames of reference}, i.e., \emph{equations are relations between (general) tensors}. 
}
\end{quote}

Let's assume we are interested in writing equations for some scalar field $\phi$. Then, our theory would be generally covariant if given a spacetime $\spacetime$ -- which we assume to be a solution to Einstein's field equations –– the equations of motion (EOM) of the scalar \textit{take the same form in any coordinate system}. For example, the equation $\phi_{;a}^{\,\,\,\, ;a} = 0$ is generally covariant, while $\phi_{,a}^{\,\,\,\, ,a} = 0$ is not.

Now, since a change of coordinates is a diffeomorphism from $\manifold$ to itself, we could say that the EOM of the field are  \emph{diffeomorphism invariant}. This is true if we regard the diffeomorphism as a \emph{passive} one, \emph{i.e.}, such that all it does is `permute' the points of the manifold. In other words: giving different names to points should not affect the physics.\\
There is nothing deep in this general covariance feature of equations. One can `covariantize' everything in Minkowski spacetime to make equations look `GR-like' without any effort whatsoever. As stated in \cite{Norton2003-GeneralCovariance&Gauge}, Kretschmann even claimed that \emph{any} theory can be made generally covariant, although this might require quite a lot of mathematical ingenuity. We have not found a proof of this, though it seems to be generally accepted \cite{RovelliGaulLQG&Diffeomorphism2000} that this is the case.\\

\emph{Active} diffeomorphisms are quite different, however, and not every theory is invariant under their action. When people say that GR is a \emph{diffeomorphism invariant} theory, what they really mean is that GR is invariant under \emph{active} diffeomorphisms. This happens to have profound implications in the interpretation of spacetime, as opposed to the `triviality' of general covariance, which is physically meaningless.\\
Perhaps the source of confusion stems from the fact that active and passive diffeomorphisms are merely different interpretations of a single mathematical transformation (a diffeomorphism). For instance, a diffeomorphism locally given by smooth maps
\begin{equation}\label{eq:diffeo_local}
f: x^a\longrightarrow f^a(x)
\end{equation}
produces a new metric $\tilde{g}_{ab}$ from an old metric $g_{ab}$ by means of the usual chage of coordinates formula:
\begin{equation}
\tilde{g}_{ab} (x):= \frac{\partial f^c(x)}{\partial x^a}\frac{\partial f^d(x)}{\partial x^b}g_{cd}(f(x)).
\end{equation} 
However, there is nothing in this formula telling us whether we should interpret the diffeomorphism of Eq.~\eqref{eq:diffeo_local} as a passive or as an active one. A mathematical formula \emph{per se} does not `mean' anything: it is pure structure~\footnote{Although in Tegmark's `mathematical universe' \cite{Tegmark:2007ud} a formula can actually mean `everything', including the interpretation (`baggage') that we humans give to the formula itself. In this hypothesis, ``\emph{\ldots{}our successful [physical] theories are not mathematics approximating physics, but mathematics approximating mathematics}''. We shall not worry about this, even when in principle we endow the \emph{External Reality Hypothesis}, that Tegmark claims to imply by necessity the \emph{Mathematical Universe Hypothesis}.}.\\
At this point, to see the difference between passive and active diffeomorphisms, it is convenient to switch to a coordinate-free formulation. In order to do so, let us first define a metric $\metric$ in such a coordinate-free way~\footnote{There are more rigorous and abstract definitions in terms of fiber products and tangent bundles. For us, this version will suffice.}:
\begin{defn}
A \emph{metric} $\metric$ is a map from points $p\in\manifold$ to the tensor product of the cotangent space at those points, \emph{i.e.}, 
\begin{align}
\metric: \, &\manifold \rightarrow {TM}^{*}\otimes{TM}^{*}\\
&p \longmapsto \metric(p).
\end{align}
\end{defn}
By Sylvester's law of inertia~\footnote{Essentially, a theorem stating that the number of pluses and minuses in a quadratic metric  depends only on whether you come from high energy physics, or GR.} we can write:
\begin{equation}
\metric = -\bm{\Theta}_0\otimes\bm{\Theta}_0 + \sum_{i=1}^{3}\bm{\Theta}_i\otimes\bm{\Theta}_i,
\end{equation}
where $\bm{\Theta}_k$ ($k=0,\ldots, 3$) form a basis of one-forms (the subindex is merely a label, \emph{not} a covariant index). Then, if $\vec{E}_l$ ($l=0,\ldots, 3$) are the dual basis of vector fields, we can define a contraction of the metric in such a way that it gives a \emph{function} $s(p)$ as an output~\footnote{We omit the details. The idea is that the contraction should be defined in such a way that each one-form gets paired with one dual vector field. Each pairing produces a function. The tensor product of functions becomes trivially equivalent to simple multiplication of functions and, since a sum of smooth functions is a smooth function, we get the desired result.}. This function is smooth and it can be used to define a notion of distance $d_{\metric}(p,p^\prime)$ between any two points $p$ and $p^\prime$ on the manifold by integration of the one-form $\mathrm{d} s$:
\begin{equation}
d_{\metric}(p,p^\prime)\equiv \int_{p}^{p^\prime}\mathrm{d} s.
\end{equation} 
It is in this sense that we can say that the metric defines a map from the Cartesian product of $\manifold$ with itself to the real set:
\begin{align}
d_{\metric}: \, &\manifold\times\manifold \rightarrow \mathbb{R}\\
&(p, p^\prime) \longmapsto d_{\metric}(p, p^\prime).
\end{align}

Take $\metric$  to be a solution to Einstein's field equations (EFE) $\vec{G} = 8\pi \vec{T}$. Then $\spacetime$ is a valid spacetime and every classical equation of motion of test particles will have a unique solution in some region, provided that this region can be foliated into spacelike hypersurfaces. It is clear that a passive diffeomorphism gives the same physical situation, because the way we label the points of $\manifold$ is not associated to any physical observable~\footnote{Although some labellings are more sensible than others and can make our lives much easier. Also, two different coordinate systems related by a diffeomorphism may have `problems' around certain points. In those cases, we have to resort to \emph{local} diffeomorphisms. (\emph{e.g.}, the transition from Cartesian to polar coordinates is not diffeomorphic when we are close to the origin.)}.\\
Now, consider a diffeomorphism $\Phi$ from $\manifold$ to itself. In general, we have
\begin{equation}
d_{\metric}(p,p^\prime) \neq d_{\metric}(\Phi^{-1}(p),\Phi^{-1}(p^\prime)).
\end{equation}

\begin{figure}
\begin{center}
\includegraphics[width=0.46 \textwidth]{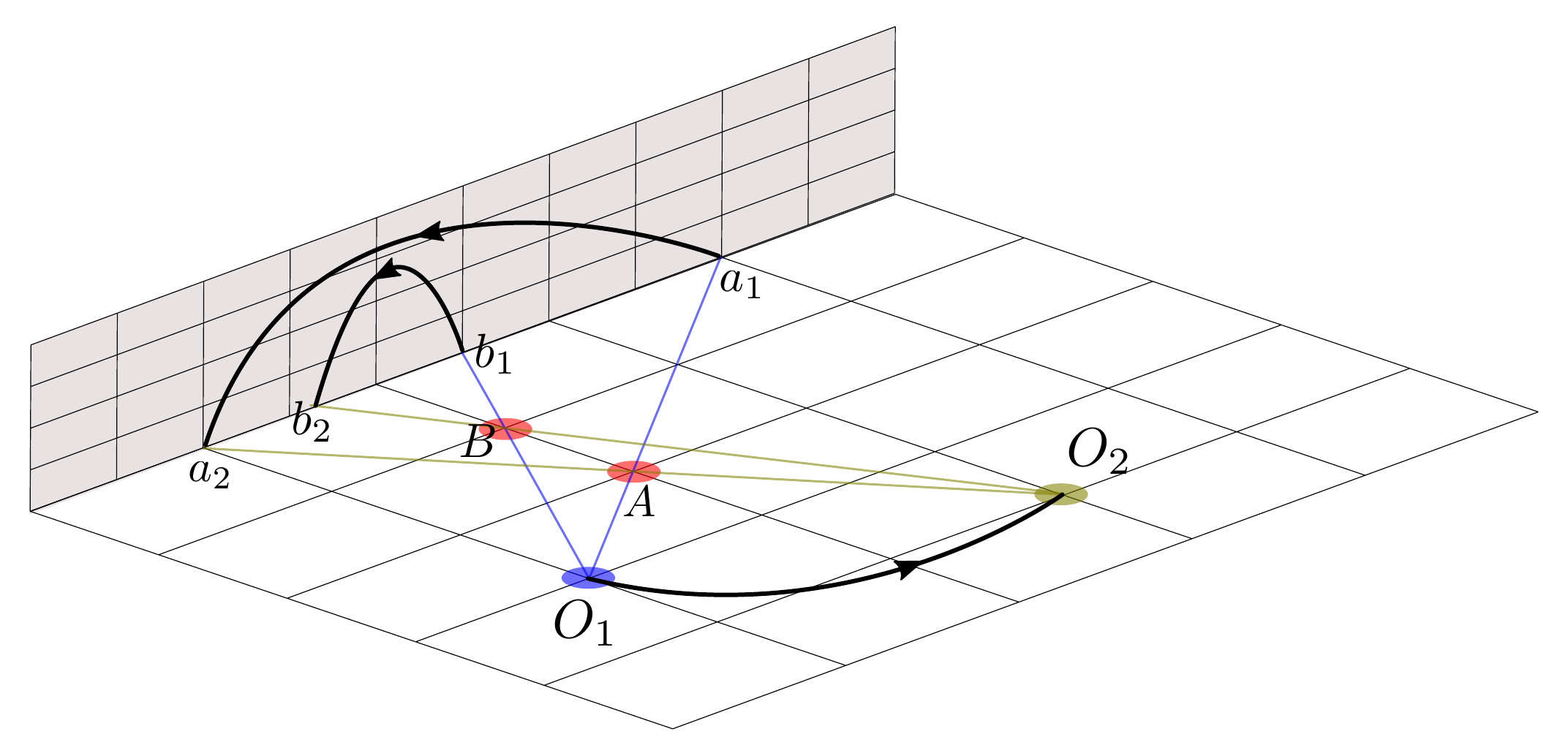}
\caption[Passive diffeomorphism]{A passive diffeomorphism, acting on a static space for simplicity, just changes our point of view: Observer $O_1$ uses an artificial background to label events $A$ and $B$ --simply positions of objects-- as $a_1$ anb $b_1$. Observer $O_2$ does the same, obtaining $a_2\neq a_1$ and $b_2\neq b_1$, respectively. The passive diffeomorphism is an invertible map that relates two descriptions of the same situation, and its effect is pictured as a set of black arrows. The white grid in which `things happen' is taken to represent a `two-dimensional ruler' for the spacetime $\manifold$, \emph{i.e.}, a metric. The artificial background can be thought as the `language' and the passive diffeomorphism as the `dictionary'.}\label{fig:passive}
\end{center}
\end{figure}

However, since $\Phi^{-1}$ is a smooth map, it ought to be possible to define a \emph{new} metric $\tilde{\metric}$ on $\manifold$ such that its associated distance function is given by
\begin{equation}
d_{\tilde{\metric}}(p,p^\prime)\equiv d_{\metric}(\Phi^{-1}(p),\Phi^{-1}(p^\prime)).
\end{equation}
The pair $(\manifold, \tilde{\metric})$ is still a valid spacetime and the claim is that it is physically indistinguishable from $\spacetime$. This is equivalent to saying that $\tilde{\metric}$ solves the \emph{same} equations (Einstein's) as $\metric$ does. This is what the sentence ``GR is a diffeomorphism invariant theory'' really means and it is clearly not a trivial fact (as passive diffeomorphism invariance is). Invariance under passive diffeomorphisms talks about the \emph{form} of equations. Invariance under active diffeomorphisms tells us something about the \emph{mathematical structure} of the theory and inevitably hints towards a \emph{relational}~\footnote{In \cite[Sec. 6.2.]{rickles} it is argued that the correct statement is that GR suggests a \emph{structuralist} view rather than a \emph{relational} one: ``\emph{(\ldots) that the observables are indifferent to matters of spacetime point role does not imply there are no spacetime points}''. For us, the word `relation' is much more suggestive than `structure' in the sense that we aim to picture the physical world as a network of information being exchanged between observers. Yet, this is not the place to discuss `observers' and `information': that would inevitably lead us to adopt an interpretation of the density operator in rigged Hilbert spaces.}  interpretation: as long as the relations between entities remain the same, it does not matter `where' or `how' they are distributed.\\
To put it in an intuitive, physical way: imagine a two-dimensional grid with objects on it. For simplicity, assume that time is `frozen' and that the picture is static (see Fig. \ref{fig:passive}). We are standing somewhere \emph{inside} the grid and looking at things. When we change our position with respect to the grid, the objects will appear to move with respect to each other but they will remain fixed to their position on the grid, because if we move with respect to the grid we can always say that it's the grid that moved with respect to us (if there were such an absolute frame of reference).  This is a passive diffeomorphism: we haven't really changed anything, just our point of view. Thus, we shouldn't expect to get a different set of equations of motion.\\

\begin{figure}
\begin{center}
\includegraphics[width=0.46 \textwidth]{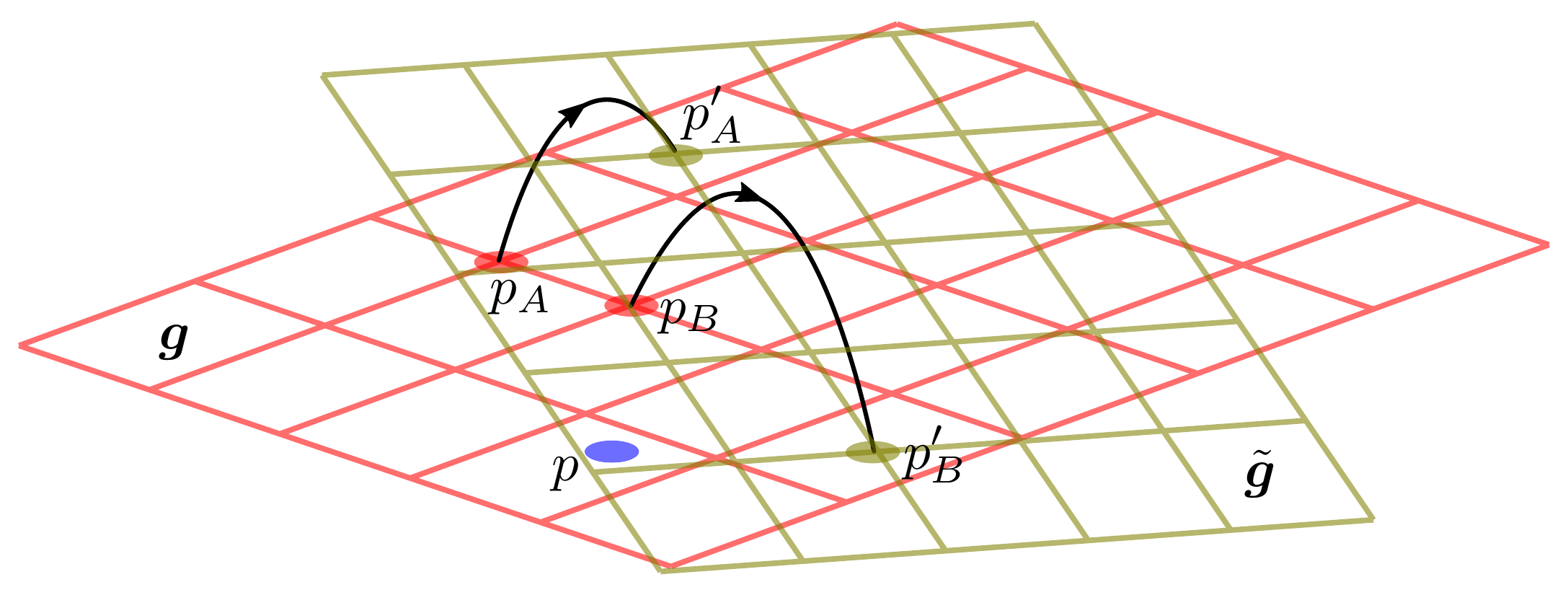}
\caption[Active diffeomorphism as seen by internal observer]{An active diffeomorphism acting on a solution to EFE produces another solution: Observer at point $p$ describes the relations between points $p_A$ and $p_B$ by means of a metric $\metric$ pictured as a red grid. An active diffeomorphism \emph{moves} the points to $p^\prime_A$ and $p^\prime_B$. Then, there exist a \emph{different} metric $\tilde{\metric}$ (green lattice) such that the relations between the new points are physically indistinguishable from the previous case. Observer $p$ has then the freedom to choose which description to use.}
\label{fig:active}
\end{center}
\end{figure}

Now imagine that we stay fixed to our position in the grid but the objects move in a smooth but otherwise arbitrary way with respect to it, and --consequently-- with respect to us. Then it is not so trivial to `see' that the physical phenomena predicted by GR remains unchanged, \emph{i.e.}, to see that it is possible to find a new grid (a new metric) which, used as a measuring tool for the new distribution of the objects (the `energy-matter'), will give us the same exact physics as we had before.\\
Let us illustrate this ideas with two complementary figures of an active diffeomorphism. In Fig. \ref{fig:active} we emphasized the active part of the diffeomorphism, in the sense that if does move the matter `around'. But even if some sort of diabolical $n$-dimensional creature (with $n\geq5$) performed (instantaneous) rearrangements of the matter in the universe we wouldn't be able to tell from the inside~\footnote{As far as GR is concerned, of course.}. In a sense, that evil creature would be mapping our entire 4-dimensional universe to a different, but physically indistinguishable one, so she would see something like what is shown in Fig. \ref{fig:active}.

\begin{figure}[h]
\begin{center}
\includegraphics[width=0.46 \textwidth]{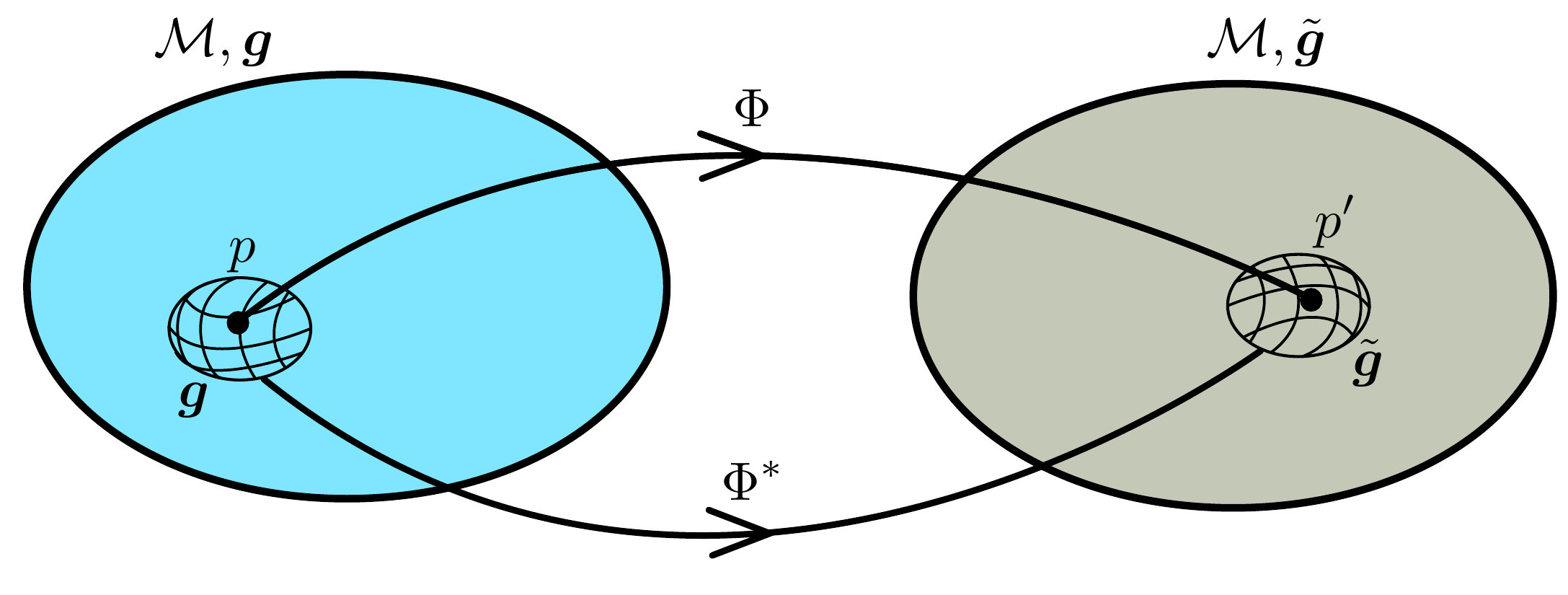}
\caption[Diffeomorphism between two spacetimes]{The action of the diffeomorphism $\Phi:\manifold\rightarrow\manifold$ on the spacetime $\spacetime$ produces a new spacetime $(\manifold, \tilde{\metric})$ that is physically equivalent to the former. A point $p\in\manifold$ gets moved to $\Phi(p)=p^\prime\in\manifold$ and the metric, represented as a grid, gets `moved' to a new one, $\tilde{\metric}$ under the action of $\Phi^{*}$, a map induced from $\Phi$ that maps metrics to metrics. This map is the pull-back and it is an isomorphism between products of cotangent spaces at points $p$ and $\Phi(p)$.}\label{fig:active2}
\end{center}
\end{figure}

In the language of gauge theories, it is often stated that $G\equiv\text{Diff}(\manifold)$ is the \emph{gauge group} of GR. This means that there is some sort of `redundancy' in the theory (or extra degrees of freedom), since two mathematically different solutions $\spacetime$ and $(\manifold,\tilde{\metric})$ are considered physically indistinguishable if they are related to each other by an (active) element of $G$. Gauging this group, \emph{i.e.}, making it \emph{local}, leads to GR. Then, once is `forced' to write down equations that  are gauge invariant, because the choice of a gauge is completely arbitrary. The same thing happens in electrodynamics, as we learn when we study quantum field theory: when one writes down the Lagrangian for a \emph{free} electron (in flat spacetime) and imposes --based on common sense-- that the physics (that is, the Lagrangian) remains invariant under \emph{local} phase shifts (represented by elements of the group $U(1)$), then one naturally picks up a new vector field $\vec{A}$, that also has to transform in a nice way and that we call the `gauge field'. The free field then couples to the gauge field, producing an \emph{interaction} term in the Lagrangian that is gauge-invariant. Neither the original electron field, nor the gauge field are gauge invariant, but the way they appear in the Lagrangian is through a gauge invariant term. One then proceeds to define gauge invariant things like the tensor field $\vec{F}:=\mathrm{d} \vec{A}$ and claims that the physics of the theory is contained \emph{only} in those objects.\\
Likewise, diffeomorphisms in GR are regarded as extra, unphysical degrees of freedom: the physics must be contained only in gauge-invariant quantities. This is in flagrant contrast with what experience tells us: in `real life' things are constrained to fixed frames of reference, and one can measure `gauge-variant' \cite{Rovelli:2013fga} quantities, such as the energy, proper time, the electric field, and so on.\\
Systems \emph{couple} to each other through certain fields, which are gauge-variant. By means of these couplings, `relative observables' appear in the equations. In \cite{Rovelli:2013fga}, the author claims that this leads to a \emph{relational} interpretation:
\begin{quotation}
``\textit{The fact that the world is well described by gauge theories
expresses the fact that the quantities we deal with
in the world are generally quantities that pertain to relations
between different parts of the world, that is, which
are defined across subsystems.}''
\end{quotation}

\section{An example}
Take for instance two isolated observers, Felice and Dob, moving along world lines $x_F^a(\tau_F)$ and $x_D^a(\tau_D)$, parametrised by their proper times. Between them there is a region in which there is a gravitational field described by some metric $\metric$. \emph{We}~\footnote{We italize for obvious reasons: we have more information than the total information of the two \emph{isolated} observers: we know the metric! In Ref.~\cite{Anderson:1995tt} -- where the author discusses the `problem of time', the dichotomy between the notion of time as seen by us, `god-like' observers and that based on the information available to the true observers is achieved by the consistent use of \emph{theorists' time} and \emph{participant-observers' time}. Likewise, in Ref.~\cite{Tegmark:2007ud}, the author uses the terms `bird' and `frog'. In the philosophy of physics literature, this vantage point of view is often referred to as `Archimedean point', of \emph{punctus Archimedis}. \label{ftn:birds_and_frogs}} `know' this, but they don't: they are isolated from the rest of the world in small elevators without windows. Each observer uses a clock to measure their proper time. We know how to write this: it is the line integral of the gravitational field along a single world line. For Felice we have that the time elapsed between events $P,Q\in x_F^a(\tau_F)$ is
\begin{equation}\label{eq:gauge_invariant_time}
\Delta\tau_{F}(P,Q) = \int_{P}^{Q} \diff \tau_F \sqrt{-g_{ab}(x_F(\tau_F))\frac{\diff x_F^a(\tau_F)}{\diff \tau_F}\frac{\diff x_F^b(\tau_F)}{\diff \tau_F}}.
\end{equation} 
The last expression is obviously gauge-invariant, but it doesn't correspond to what Felice would write. How would she know what values to assign to the events $P,Q$ if she only has access to her small, sad elevator? Instead, she would simply get a number by looking at a clock: something that she would write as
\begin{equation}\label{eq:gauge_variant_time}
\Delta\tau_F= \int_0^1\diff \tau_F\sqrt{-g_{00}(\tau_F,0,0,0)},
\end{equation}
where the integration limits denote the same events $P,Q$ but as described by Felice: when the clock hand is here and when it is there. We set these to be 0 and 1 with no loss of generality.  For her, there is no way of knowing the function $g_{00}$ for arbitrary values of the arguments. Through local experiments, all she can do is erect a local coordinate system (a tetrad) and assume that there might be a metric $g_{ab}$ out there. Then, that hypothesised metric should enter her description of time in the way of equation Eq.~\eqref{eq:gauge_variant_time} above. This number $\Delta\tau_F$ is clearly gauge-dependent as it involves a component of the metric tensor in a particular frame of reference. The same thing would happen to Dob. Let us assume that he is also measuring the time lapsed between events $P$ and $Q$. In other words, the world lines of Felice and Dob meet at these points. They don't have to collide: passing each other by a few meters will do.\\
For Dob, things are completely similar: he measures some gauge-variant number $\Delta\tau_D$.

It is only when the two observers exchange information \emph{through} the causal structure given by the gravitational field --\emph{i.e.}, by \emph{coupling} to the field-- that they can describe the situation with the gauge-invariant quantity of equation Eq.~\eqref{eq:gauge_invariant_time}. In other words, based on the knowledge of $g_{ab}$, they can find and predict the gauge-invariant  function $\tau_F(\tau_D)$ (or its inverse $\tau_D(\tau_F)$).\\
Gauge-variant quantities can be used to construct gauge-invariant ones by promoting the original system of two isolated \emph{local} observers to a larger system given by them \emph{and} the field. In practice, this promotion is effectively achieved by the exchange of information between the two observers. In many situations, this exchange of information --signalling-- involves \emph{classical} physics, as in the following example (borrowed from \cite{Rovelli:2013fga}): an observer on Earth with two similar clocks throws one of them up in the air (proper time $\tau^\prime$) and, when it comes back, he compares the value of $\tau^\prime$ with $\tau$ (proper time on Earth). GR correctly predicts the function $\tau(\tau^\prime)$ through the knowledge of $g_{ab}$. GR cannot predict $\tau$ or $\tau^\prime$ alone, although both of them are measurable. Hence the \emph{relational} nature of gravity.\\
In Fig. \ref{fig:active} above we already saw this from the mathematical point of view: diffeomorphism invariance of EFE means that only relations between events can be predicted.\\
Let us quote some examples that Rovelli \cite{Rovelli:1999hz} gives as physical gauge-invariant quantities that can be predicted by GR  and confirmed experimentally:
\begin{quote}
{\setstretch{1.0}
``Examples of diff-invariant quantities (\ldots) are the Earth-Venus distance during the last solar eclipses, the number of pulses of a pulsar in a binary system that reach the Earth during one revolution of the system (that is, between two Doppler maxima), the energy deposited on a gravitational antenna by a gravitational wave pulse and, in fact, any significative physical
quantity measured in general relativistic experimental or observational physics.''
}
\end{quote}
When the information exchanged behaves quantum mechanically, things become far more subtle, as we should expect. This is because in most of the interpretations of quantum mechanics (QM), the world splits into two: the system under observation (SUO) and the observer (O), which is linked to an apparatus. We shall now adopt the terminology of \cite{Anderson:1995tt}: By `theorists' we will mean \emph{us}, \emph{i.e.}, beings with access to the whole theory (what we have called `god-like observers'); and by `participant-observers' we will mean the real observers, that describe the world around them collecting information by means of experiments.
\\
From the theorist's point of view, in GR is possible to enlarge systems so that, through the coupling of $g_{ab}$ to the original system, gauge-invariant quantities involving gauge-variant objects can be calculated. The calculation of gauge-invariant quantities means that \emph{predictions} can be made. There is, in principle, nothing wrong with this, for if the assumptions made by the theorist were wrong, experiment will tell. Thus, a theorist can couple distant objects in the universe by means of the theory of GR, make predictions, and then ask their experimental colleagues to test them. In a way, the role of the theorist is completely irrelevant insofar as the outcomes are concerned: there is only one way things can happen in a `block universe'. This is a clear consequence of the classicality of GR: the observer does not affect the SUO, which evolves in a predetermined way regardless of what `questions' (\emph{i.e.}, measurements) are asked.

We shall disregard the problem of the `conscious observer'. In principle, an observer could be `anything'. What makes humans different from, say, turtles, is that the former can perform \emph{experiments}, recording data from measurements in notebooks; whereas the latter, being an `unconscious' system, may not have a memory and no notebook available. One could say that any interaction represents a measurement. But it is clear that a collection of measurements is not an experiment. Let us close this rather philosophical paper with some words by Anderson \cite{Anderson:1995tt}:
\begin{quote}
{\setstretch{1.0}
 ``\textit{There is a tendency to anthropomorphize observers which
associates with them a connotation of awareness. This connotation is wholly
undeserved, and unnecessarily complicates the language of measurement and
observation. I shall speak of observers even in the absence of human beings.
An observer is simply a subsystem whose state we choose to focus on as it
interacts with other subsystems.}''
}
\end{quote}

\section*{Acknowledgments}
I thank George Jaroszkiewicz, my advisor while at Nottingham.

\end{document}